\documentclass[english,10pt,final,twocolumn]{elsarticle}
\usepackage[T1]{fontenc}
\usepackage[latin9]{inputenc}
\usepackage{pdfpages}
\usepackage{amstext}
\usepackage{amsthm,amsmath,amsfonts,amssymb,mathrsfs}
\usepackage{graphicx}
\usepackage{gensymb}
\usepackage{placeins}
\makeatletter

\newcommand{\lyxmathsym}[1]{\ifmmode\begingroup\def\b@ld{bold}
  \text{\ifx\math@version\b@ld\bfseries\fi#1}\endgroup\else#1\fi}
\usepackage{braket}
\usepackage{mathrsfs}
\usepackage{slashed}
\usepackage{multirow}
\usepackage{bm}
\usepackage{datetime}
\usepackage{siunitx}

\DeclareSIUnit[number-unit-product = {}]\clight{c}
\DeclareSIUnit\eVperc{\eV\per\clight}
\DeclareSIUnit\GeVpercs{\giga\eV\squared\per\clight\squared}
\DeclareSIUnit\MeVpercs{\mega\eV\per\clight\squared}
\journal{Physics Letters B}
\biboptions{comma,numbers,compress}
\usepackage[left=1.6cm,right=1.04cm,top=1.65cm,bottom=1.65cm,columnsep=25pt]{geometry}
\usepackage{hyperref}  
\hypersetup{breaklinks = true, colorlinks = true, allcolors = magenta}
\usepackage{setspace}
\usepackage{graphicx}
\usepackage{wrapfig}

\makeatother
\usepackage{babel}
\begin{document}

\begin{frontmatter}{}

\title{
Comparison of recoil polarization in the $^{12}{\rm C}(\vec{e},{e}'\vec{p})$\\ process for protons extracted from $s$ and $p$ shells
}

\author[JSI]{T.~Kolar\corref{cor2}}

\ead{tim.kolar@ijs.si}

\author[TAU]{S.J.~Paul}
\author[JSI]{T.~Brecelj}
\author[Mainz]{P.~Achenbach} 
\author[Mainz]{R.~B\"ohm}
\author[zagreb]{D.~Bosnar} 
\author[Rutgers]{E.~Cline}
\author[TAU]{E.O.~Cohen}
\author[Mainz]{M.O.~Distler}
\author[Mainz]{A.~Esser}  
\author[Rutgers]{R.~Gilman}
\author[Pavia,INFN]{C.~Giusti}
\author[Mainz]{M.~Hoek}   
\author[TAU]{D.~Izraeli}  
\author[Mainz]{S.~Kegel}  
\author[Mainz]{Y.~Kohl}   
\author[NRCN,TAU]{I.~Korover}  
\author[TAU]{J.~Lichtenstadt}
\author[TAU,soreq]{I.~Mardor}  
\author[Mainz]{H.~Merkel}  
\author[JSI,Mainz,UL]{M.~Mihovilovi\v{c} }  
\author[Mainz]{J.~M\"uller}   
\author[Mainz]{U.~M\"uller}  
\author[TAU]{M.~Olivenboim}
\author[TAU]{E.~Piasetzky}
\author[Mainz]{J.~Pochodzalla}  
\author[huji]{G.~Ron}
\author[Mainz]{B.S.~Schlimme}  
\author[Mainz]{C.~Sfienti}
\author[UL,JSI]{S.~\v{S}irca}  
\author[Mainz]{R.~Spreckels}   
\author[JSI]{S.~\v{S}tajner }  
\author[USK]{S.~Strauch}
\author[Mainz]{M.~Thiel}  
\author[Mainz]{A.~Weber} 
\author[TAU]{I.~Yaron}   

\author{\\(A1 Collaboration)}

\cortext[cor2]{Corresponding author}

\address[JSI]{Jo\v{z}ef Stefan Institute, 1000 Ljubljana, Slovenia}

\address[TAU]{School of Physics and Astronomy, Tel Aviv University, Tel Aviv 69978,
Israel}

\address[Mainz]{Institut f\"ur Kernphysik, Johannes Gutenberg-Universit\"at, 55099
Mainz, Germany}

\address[zagreb]{Department of Physics, Faculty of Science, University of Zagreb, HR-10000
Zagreb, Croatia}

\address[Rutgers]{Rutgers, The State University of New Jersey, Piscataway, NJ 08855, USA}

\address[Pavia]{Dipartimento di Fisica, Universit\`{a} degli Studi di Pavia, 27100 Pavia, Italy}

\address[INFN]{Istituto Nazionale di Fisica Nucleare sezione di Pavia, 27100 Pavia, Italy}

\address[NRCN]{Department of Physics, NRCN, P.O. Box 9001, Beer-Sheva 84190, Israel}


\address[soreq]{Soreq NRC, Yavne 81800, Israel}

\address[UL]{Faculty of Mathematics and Physics, University of Ljubljana, 1000
Ljubljana, Slovenia}

\address[huji]{Racah Institute of Physics, Hebrew University of Jerusalem, Jerusalem 91904, Israel}

\address[USK]{University of South Carolina, Columbia, SC 29208, USA}

\begin{abstract}
We present the first measurements of the double ratio of the polarization-transfer components   $(P^{\prime}_{\!x} \!/ P^{\prime}_{\!z} )_p/ (P^{\prime}_{\!x} \!/ P^{\prime}_{\!z} )_s$ for knock-out protons from the $s$ and $p$ shells in $^{12}{\rm C}$ measured by the $^{12}{\rm C}(\vec{e},{e}'\vec{p}\,)$ reaction in quasi-elastic kinematics. The data are compared to theoretical predictions in the relativistic distorted-wave impulse approximation. Our results show that the differences between $s$- and $p$-shell protons, observed when compared at the same initial momentum (missing momentum), largely disappear when the comparison is done at the same proton virtuality. We observe no difference in medium modifications between protons from the $s$ and $p$ shells with the same virtuality in spite of the large differences in the respective nuclear densities.

\end{abstract}
\date{\today}

\end{frontmatter}{}

\section{Introduction}
The effects of the nuclear medium on the structure of bound nucleons and their dependence on
the average nuclear density have been subject to extensive theoretical and experimental investigations \cite{Ron:2013, Meucci:2001qc, Giusti:1989, Paul:2019, Izraeli:2018, Brecelj:2020, Strauch:2003, Boffi:1996, Noble:1981,  Schiavilla:1993, Jourdan:1995, Lu:1998, Thomas:2000, Ryckebusch:2003, Aubert:1983, Steenhoven:1987, Eyl:1995, Milbrath:1998, Malov:2000, Dieterich:2001, Hu:2006, Paolone:2010, Yaron:2017, IzraeliYaron:2018}. The $^{12}{\rm C}$ nucleus is a very appealing target to study nuclear density-dependent differences in bound nucleons. Its structure is well understood, and nuclear medium effects on the bound proton were studied by both unpolarized and polarized $(e,e'p)$ reactions \cite{Steenhoven:1986, Izraeli:2018, Brecelj:2020}. The average local nuclear density in its $s$ and $p$ shells differs by about a factor of two \cite{Ron:2013}. Hence, studying quasi-elastic processes on protons, which are sensitive to the proton form factors, should be a good approach to observe any density dependence arising from the differences between the protons extracted from the two shells. Reliable theoretical calculations for this nucleus \cite{Meucci:2001qc, Giusti:1989} facilitate the interpretation of the experimental observations.

The free nucleon structure is characterized by its electromagnetic form factors $G_{\rm E}$ and $G_{\rm M}$. In the one-photon-exchange approximation, the ratio between the transverse ($x$) and longitudinal ($z$) polarization-transfer components, $P'_x\!/P'_z$, measured by polarized elastic electron scattering, is proportional to $G_{\rm E}/G_{\rm M}$ \cite{Akhiezer:1974}. In quasi-elastic ${\rm A}(\vec{ e},{ e}'\vec{ p}\,)$ reactions, the sensitivity of $P'_x\!/P'_z$ to the $G_{\rm E}/G_{\rm M}$ ratio persists and, hence, the measurement of polarization transfer to the knocked-out proton has been suggested as a tool to investigate nuclear-medium modifications of the bound proton \cite{Glashausser:1989}. 

Theoretical calculations suggest that comparing the polarization transfer to protons knocked out from the $s$ and $p$ shells should result in measurable differences in the ratio of the polarization-transfer components \cite{Ron:2013}. We study the double ratio $(P^{\prime}_{\!x} \!/ P^{\prime}_{\!z} )_p/ (P^{\prime}_{\!x} \!/ P^{\prime}_{\!z} )_s$, which is sensitive to the deviation of the form-factor ratio, $G_{\rm E}/G_{\rm M}$, in each shell. We note that this is equivalent to $(P^{\prime s}_{\!z} \!/ P^{\prime p}_{\!z} )/ (P^{\prime s}_{\!x} \!/ P^{\prime p}_{\!x} )$ where, based on calculations discussed below, one may expect that the differences in final-state interactions (FSI) for protons knocked out from the $s$ and $p$ shells (as well as between the longitudinal and transverse components) will largely cancel. 

Almost all theoretical calculations characterize the bound nucleons by their initial internal momentum which, in the absence of FSI, is equivalent to the measured missing momentum in the reaction. However, it has been shown \cite{Izraeli:2018, Yaron:2017, Paul:2019} that deviations of the ratio $P'_x\!/P'_z$ obtained through the quasi-free reaction from that of the free nucleon, as a function of the bound-proton virtuality (see Eq.~(\ref{eq:virt})), are in good overall agreement between different nuclei, and at different momentum transfers and kinematics. This suggests that the nucleon's virtuality, which is a measure of its ``off-shellness'', might be a suitable variable to characterize the bound nucleon. Since virtuality depends also on the nucleon binding energy, nucleons from different shells with the same missing momentum do not have the exact same virtuality. However, we chose the kinematics for the measurements so that there is an overlap between $s$- and $p$-shell removal for both the missing momentum and the virtuality.

We present here the measurements of polarization transfer to the protons extracted from $s$ and $p$ shells in $^{12}\mathrm{C}$ in search of nuclear-density-dependent modifications of the bound protons.
We study the transverse-to-longitudinal components ratio, and compare the results from the two shells by the aforementioned double ratios. The data are also compared to calculations in the relativistic distorted-wave impulse approximation (RDWIA) \cite{Meucci:2001qc} which use free-nucleon electromagnetic form factors. We present the comparison in both missing momentum and bound-proton virtuality, and demonstrate the advantage of using the latter as a parameter for such comparisons.
\section{Experimental Setup and Kinematics}
\label{sec:setup}
\FloatBarrier
The experiment was carried out in the A1 Hall at the Mainz Microtron (MAMI) using a $600$ $\rm MeV$ continuous-wave (CW) polarized electron beam of about $10$ $\rm \mu A$. The measurements were performed at $Q^2=0.175$ ${\rm GeV}^2\!/c^2$. The beam polarization, $P_{\! e}$, was measured periodically using the standard M{\o}ller \cite{Wagner:1990, Bartsch:2002} and Mott \cite{Tioukine:2011} polarimeters. The polarization range was $80.5\%<P_{\! e}<88.7\%$. The polarization was increasing at the beginning of the experiment with the decrease of the quantum efficiency towards the end-of-life of the strained GaAs crystal used as the beam source. It dropped after the annealing process of the crystal. To account for the variations in $P_{\! e}$ we used a rolling average of the measurements (resetting it after the refreshing process), which was applied in the analysis of the data. 
\begin{figure}[th]
\begin{center}
\includegraphics[width=\columnwidth]{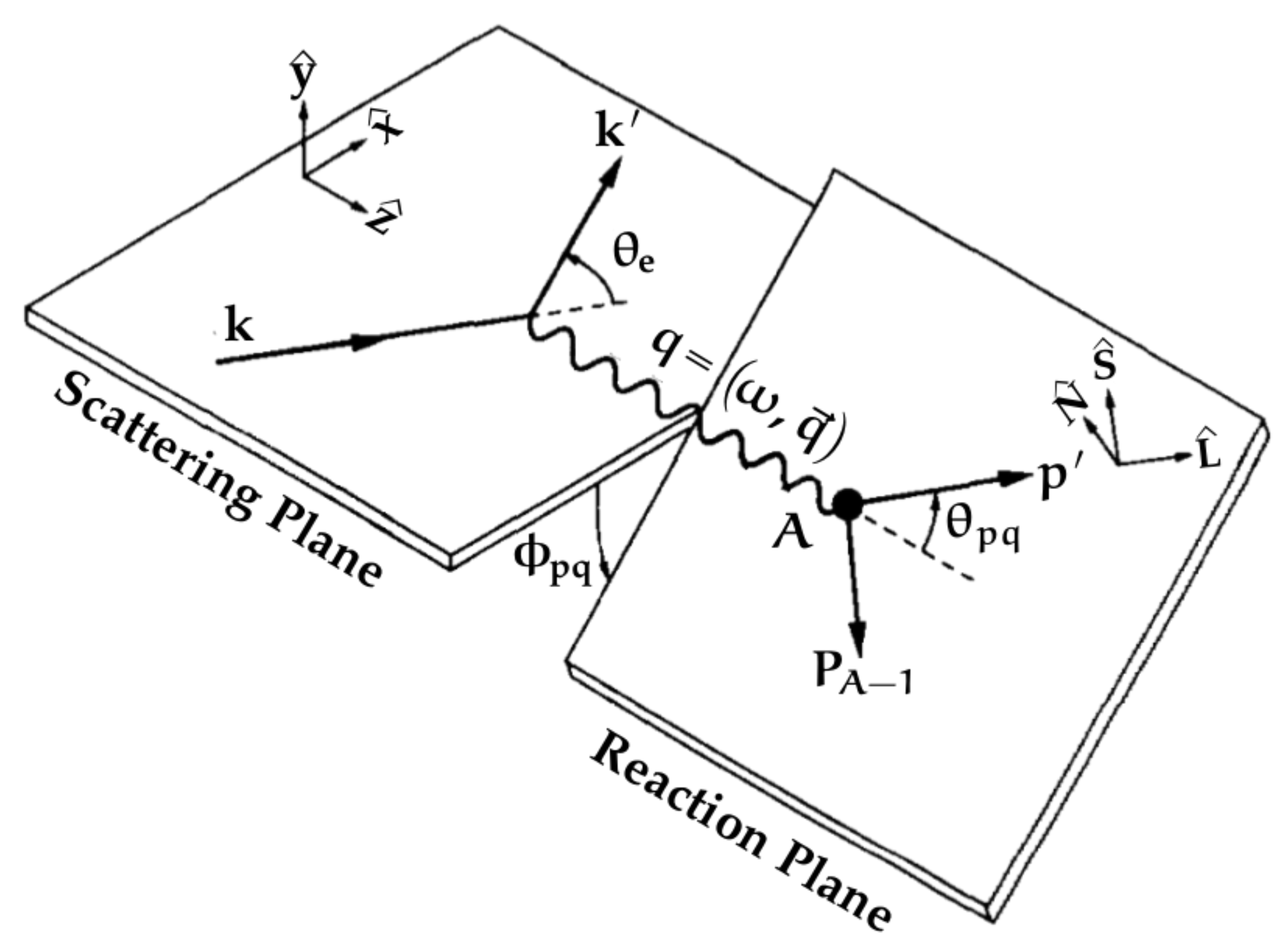}
\caption{Kinematics of the measured $\mathrm{A}(\vec{e},e'\vec{p})$ reaction. The scattering plane is spanned by the ingoing and outgoing electron momentum, $\vec{k}$ and $\vec{k}^{\prime}$, respectively. The reaction plane is spanned by the transferred momentum, $\vec{q}$, and the outgoing proton momentum, $\vec{p}^{\,\prime}$. We choose to represent the polarization components in the scattering plane by using a right-handed coordinate system with its axes being: $\hat{z}$ parallel to the momentum transfer $\vec{q}$, $\hat{y}$ along the cross product of the ingoing and outgoing-electron momentum, $\vec{k}\times\vec{k}^{\prime}$, and $\hat{x}=\hat{y}\times\hat{z}$. Another often-used reference frame is $\hat{L}\hat{N}\hat{S}$ where $\hat{L}$ points along the outgoing proton momentum, $\vec{p}^{\,\prime}$, $\hat{N}$ is along $\vec{p}^{\,\prime}\times\vec{q}$, and $\hat{S}=\hat{N}\times\hat{L}$. There are three important angles that help to characterize the reaction above. The electron scattering angle, $\theta_{e}$, together with the energy of an ingoing electron, $k^0$, determines the momentum transfer. The azimuthal angle between $\vec{q}$ and $\vec{p}^{\,\prime}$, $\phi_{pq}$, represents the angle between the scattering and reaction plane, whereas $\theta_{pq}$ is the corresponding polar angle.  }
\label{fig:kinPlanes}
\end{center}
\end{figure}
\begin{table}
\caption{
Central kinematics of the $^{12}$C($\vec{e},e'\vec p)$ data presented in this work.}
\begin{center}
\begin{tabular*}{\columnwidth}{l @{\extracolsep{10mm}} l @{\extracolsep{\fill}} r}
\hline\\[-8pt]
$E_{\rm beam}$ & [MeV] & 600 \\[1pt]
$Q^2$          & [${\rm GeV}^2\!/{c}^2$] & 0.175 \\[1pt]
$p_e$          & [${\rm MeV}\!/c$] & 368 \\[1pt]
$\theta_e$     & [$\degree$] & $-$52.9 \\[1pt]
$p_p$          & [${\rm MeV}\!/c$] & 665  \\[1pt]
$\theta_{p}$   & [$\degree$] & 37.8 \\[1pt]
$p_{\rm miss}$ & [${\rm MeV}\!/{c}$] & $-$270 to $-$100 \\[1pt]
$\nu$          & [${\rm MeV}^2\!/{c}^2$] & $-$160 to $-$40 \\[1pt]
\hline
\end{tabular*}
\end{center}
\label{tab:kinematics}
\end{table}%

We used a $^{12}{\rm C}$ target consisting of three $0.8$ mm-thick foils, which were rotated $40^{\circ}$ relative to the beam. This way we minimized the path of the outgoing proton through the target and, thus, reduced the energy loss and the probability of multiple scattering. The two high-resolution spectrometers of the A1 Hall \cite{Blomqvist:1998} were used to analyze the scattered electrons (Spectrometer C) and the knock-out protons (Spectrometer A). In Spectrometer A we installed a focal-plane polarimeter (FPP) \cite{Pospischil:2002FPP} in which the polarized protons experience secondary scattering on a carbon analyzer, resulting in an angular asymmetry due to the spin-orbit part of the nuclear force. Its angular distribution is given by
\begin{equation}\label{eq:FPPDistro}
\frac{\sigma(\vartheta,\varphi)}{\sigma_0(\vartheta)}=1+A_C(\vartheta,E_{p'})(P_y^{FPP}\cos\varphi-P_x^{FPP}\sin\varphi)\,,
\end{equation}
where $\sigma_0(\vartheta)$ is the polarization-independent part, $A_C$ is the analyzing power of the carbon scatter er, $\vartheta$ is the polar angle of secondary scattering, $\varphi$ is the azimuthal angle, and $P_x^{\rm FPP}$ and $P_y^{\rm FPP}$ are the transverse polarization components of the proton at the focal plane. The analyzing power depends on the energy of the outgoing proton $E_{p'}$ and was adopted from \cite{AprileGiboni:1984, Mcnaughton:1985}. To measure this distribution, horizontal drift chambers (HDCs) \cite{Pospischil:2002HDC} were placed behind the scatterer. 

Table \ref{tab:kinematics} and Figure \ref{fig:kinPlanes} show the kinematic setting we used and the relevant kinematic variables. We use a convention where the sign of the magnitude of the missing momentum, $\vec{p}_{\rm miss}=\vec{q}-\vec{p}^{\,\prime}$, is determined by the sign of $\vec{p}_{\rm miss}\cdot\vec{q}$. We define the virtuality of the embedded nucleon as:
\begin{equation}\label{eq:virt}
\nu\equiv\Big(m_{A}-\!\sqrt{m_{A-1}^{2}+p_{\rm miss}^{2}}\,\Big)^{2}\!-p_{\rm miss}^{2}-m_{p}^{2},
\end{equation}
where $m_p$, $m_{A}$, and $m_{A-1}\equiv\sqrt{(\omega - E_{p'}+ m_{A})^2-p^{2}_{\rm miss}}$ are the masses of the proton, target nucleus ($^{12}\mathrm{C}$) and residual nucleus ($^{11}\mathrm{B}$, not necessarily in its ground state), respectively. Here, $\omega=k^0-k^{\prime 0}$ is the energy transfer and $E_{p'}$ is the total energy of the outgoing proton. 

We chose the kinematic setting shown in Table \ref{tab:kinematics} to access protons with high missing momentum from both $s$ and $p$ shells. This corresponds to Setting B in previous measurements reported in \cite{Izraeli:2018,Brecelj:2020}. In those measurements we explored regions of positive and negative missing momenta to study the general behavior of polarization transfer and compared it between different nuclei. We now present a dedicated measurement performed in 2017 with improved statistics and a focus on a missing-momentum range where there is an overlap between the protons knocked out from the $s$ and $p$ shells in both missing momentum and virtuality. The present results were obtained from the combined data sets.

\begin{figure}[th]
\begin{center}
\includegraphics[width=\columnwidth]{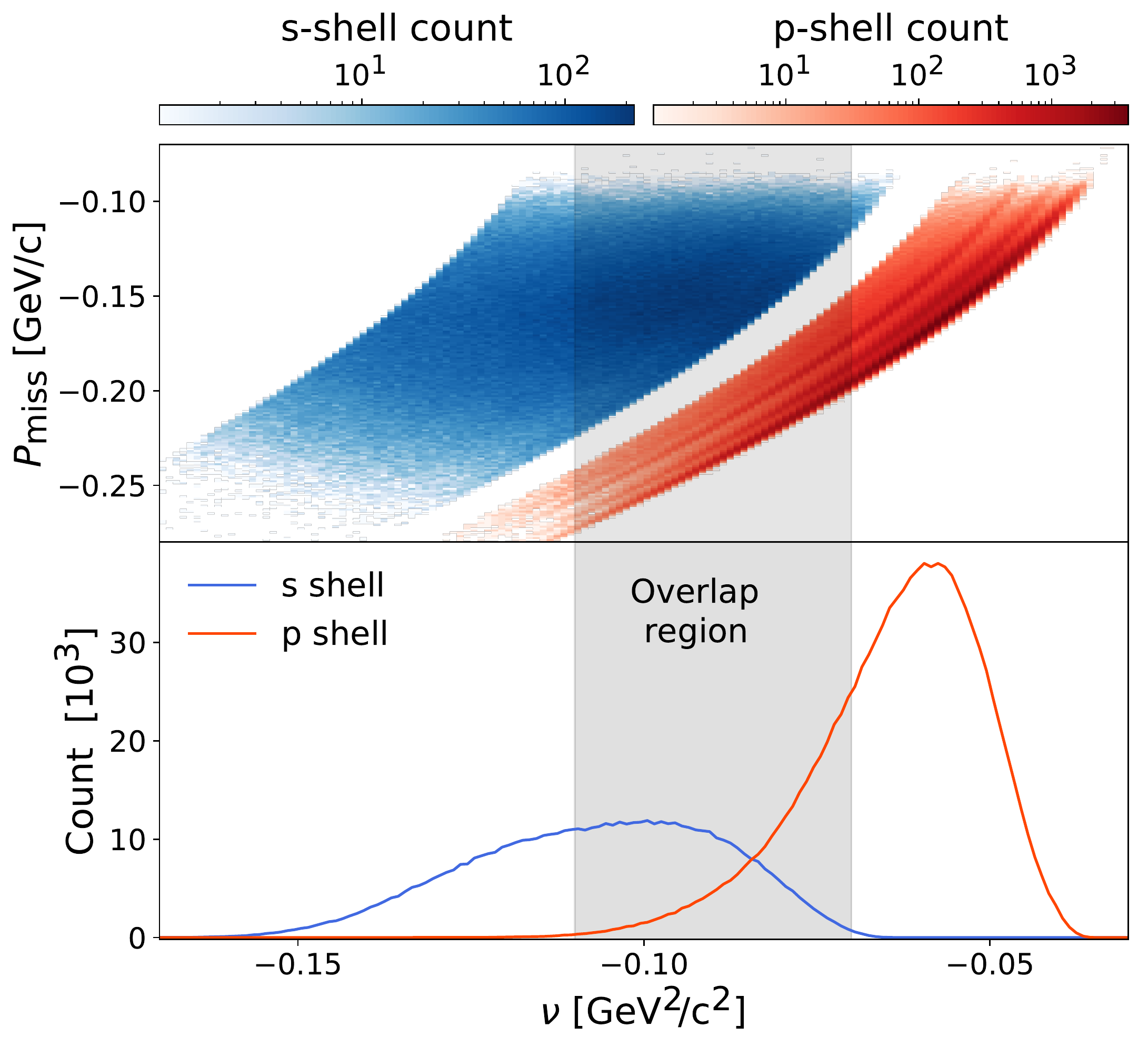}
\caption{Upper panel: the missing-momentum-versus-virtuality phase space covered by $^{12}{\rm C}(\vec{e},e'\vec{p})$ data from this experiment. Lower panel: projection of the phase-space on the virtuality axis. The gray band shows the virtuality-overlap region for protons extracted from the $s$ and $p$ shells.}
\label{fig:v-pmiss}
\end{center}
\end{figure}

We distinguished between the protons extracted from the $s$ and $p$ shells based on their measured missing energy defined as
\begin{equation}
  E_{\rm miss} = \omega - T_{p'} - T_{^{11}\mathrm{B}}\,,
\end{equation}
where $T_{p'}$ is the kinetic energy of the detected proton and $T_{^{11}\mathrm{B}}$ is the calculated kinetic energy of the recoiling $^{11}\mathrm{B}$ nucleus. Following \cite{Izraeli:2018} and \cite{Dutta:2003}, protons with $15 < E_{\rm miss} <25$ $\rm MeV$ correspond primarily to proton removal from the $p_{3/2}$ shell, while those with $30 < E_{\rm miss} <60$ $\rm MeV$ originate from the $s_{1/2}$ shell. The two step processes in the $(e,e'p)$ reaction at this kinematics are expected to be small \cite{Steenhoven:1985}. The measured missing-momentum-versus-virtuality phase space for protons from both shells obtained in our experiment is shown in Fig.~\ref{fig:v-pmiss}. The shaded area indicates the virtuality range common to both shells, and the distribution obtained from each shell is projected in the bottom panel of Fig.~\ref{fig:v-pmiss}.

\section{Determination of the Transferred Polarization and Uncertainties}
We follow the convention of \cite{Strauch:2003} to express individual components of the outgoing polarization in the scattering plane, $\boldsymbol{P}$. The coordinate system convention is shown in Fig.~\ref{fig:kinPlanes}.

To obtain the polarization components we utilized the maximum-likelihood estimation where we optimized the outgoing-proton polarization components, treated as parameters. Because our kinematics are close to parallel, we assumed only one induced component, $P_y$, \cite{Paul:2020} and two transferred components, $P_x^{\prime}$ and $P_z^{\prime}$, to be non-zero. Therefore, the total polarization of the outgoing proton at the target is
\begin{equation}\label{eq:targetPols}
\boldsymbol{P}=\left( h P_{\! e} P_{\!x}^{\prime},\, P_{\!y},\, h P_{\! e} P_{\!z}^{\prime}\right)^\mathrm{T}\,,
\end{equation}
where $h$ is the electron helicity. The contributions from other components are either very small in (anti)parallel kinematics or cancel because of their anti-symmetric dependence on the angle between the scattering and reaction planes, $\phi_{pq}$ \cite{Giusti:1989}.  

The protons travel through the magnetic fields of the spectrometer before reaching the FPP, where we measure their polarization components, $P_x^{\rm FPP}$ and $P_y^{\rm FPP}$. Therefore, before evaluating the likelihood function, we propagate the proposed estimates of target components from Eq.~(\ref{eq:targetPols}) through the spectrometer with the spin-transfer matrix $\boldsymbol{S}$ which was calculated with the \texttt{QSPIN} program \cite{Pospischil:2000}. To determine the target polarization components that best fit the measured distribution from Eq.~(\ref{eq:FPPDistro}), we maximize the log-likelihood function
\begin{equation}
\log\mathcal{L}=\sum\limits_{\rm events} \log(1+A_C(\vartheta, E_{p'})\boldsymbol{\lambda}\cdot \boldsymbol{P})\,,
\end{equation}
where
\begin{equation}
\boldsymbol{\lambda} = \left(
\begin{array}{c}
           S_{yx}\cos\varphi - S_{xx}\sin\varphi \\
           S_{yy}\cos\varphi - S_{xy}\sin\varphi \\
           S_{yz}\cos\varphi - S_{xz}\sin\varphi
\end{array}
\right)
\end{equation}
is determined per-event. It includes trajectory-dependent spin-transfer coefficients, $S_{ij}$, and the measured azimuthal angle $\varphi$ of the secondary scattering of the proton.

\begin{table}[ht]
\caption{
The sources contributing to the systematic uncertainties of the individual components, $P'_x$, $P'_z$, single ratios, $(P'_x/P'_z)_{s,p}$, and the double ratio, $(P'_x/P'_z)_s/(P'_x/P'_z)_p$.  All values are in percent.}

\begin{tabular*}{\columnwidth}{l l @{\extracolsep{\fill}} r @{\extracolsep{6pt}} r @{\extracolsep{5pt}} r @{\extracolsep{5pt}} r}
\hline\\[-9pt]
\multicolumn{2}{l}{} & $P'_x$  & $P'_z$ & $(P'_x/P'_z)_{s,p}$ & {\Large $\frac{(P'_x/P'_z)_s}{(P'_x/P'_z)_p}$} \\[7pt]
\hline
\multicolumn{2}{l}{Beam pol.}          & 2.0    & 2.0 & $\approx 0.0$    \hspace*{4mm}& $\approx 0.0$    \hspace*{3mm}\\
\multicolumn{2}{l}{Analyzing power}         & 1.0    & 1.0 & $\approx 0.0$    \hspace*{4mm}& $\approx 0.0$    \hspace*{3mm}\\
\multicolumn{2}{l}{Beam energy}             & 0.2    & 0.6 & 0.8    \hspace*{4mm}& $<$0.1 \hspace*{3mm}\\
\multicolumn{2}{l}{Central kinematics}      & 0.6    & 0.8 & 0.9    \hspace*{4mm}& 0.1    \hspace*{3mm}\\
\multicolumn{2}{l}{Alignment}               & $<0.1$ & 0.1 & 0.1    \hspace*{4mm}& 0.3    \hspace*{3mm}\\
\multicolumn{2}{l}{Software cuts}           & 1.7    & 2.1 & 1.9    \hspace*{4mm}& 0.8    \hspace*{3mm}\\
\multirow{2}*{$E_{\rm miss}$ cut} & $s$ shell  & $<0.1$ & $<0.1$ & $<0.1$    \hspace*{4mm}& \multirow{2}*{0.6}     \hspace*{3mm}\\
                                  & $p$ shell  & 0.2   & 0.5 & 0.6    \hspace*{4mm}& \\
\multicolumn{2}{l}{Precession (STM fit)}    & 0.3    & 0.3 & $<0.1$   \hspace*{4mm}& $<0.1$    \hspace*{3mm}\\
\multicolumn{2}{l}{Precession (trajectory)} & 0.2    & 0.3 & $<0.1$ \hspace*{4mm}& $<0.1$   \hspace*{3mm}
\\
\hline
\multicolumn{2}{l}{Total}                   & 3.4    & 3.7 & 2.3    \hspace*{4mm}& 1.1    \hspace*{3mm}\\
\hline
\end{tabular*}
\label{tab:systematics}
\end{table}

The uncertainties of the extracted components and their ratios were estimated through the numerical second-order partial derivative of the log-likelihood function and, besides the numerical error, include a part of the systematic spin-transfer error as well. As can be seen in Table \ref{tab:systematics}, the beam polarization and the analyzing power are the largest contributors to the uncertainty in the polarization components  $P'_x$ and $P'_z$, while their effect largely cancels when we form either a single or a double ratio. The uncertainties in the beam energy and the central kinematics affect the basis vectors of the scattering-plane coordinate system and influence the binning of the events. Another important contributor to the uncertainty when determining the secondary-scattering distribution is the quality of the alignment between the tracks extrapolated from the vertical drift chambers to the HDC plane and those measured by the HDCs themselves. 

The above three sources of uncertainty (beam energy, central kinematics, and detector alignment) were studied through the repetition of the analysis with modified values. We modified each contributor separately by its uncertainty value, and determined how much this affected the extracted polarizations. Similarly, we determined the contributions from various software cuts employed in the analysis, by placing each of them slightly tighter and looking for the average effect of the modified cut over all of the bins. Because a modification of the cut always impacts the number of considered events, we performed a parallel re-analysis, where we left the chosen cut unchanged but reduced the number of events by a random selection. 

Another possible source of the systematic uncertainty is the separation of the protons from the $s$ and $p$ shells by the missing-energy cut. Although the neighboring boundaries of the two $E_{\rm miss}$ ranges are $5$ $\mathrm{MeV}$ apart, each of them contains a small amount of protons coming from the other shell. To estimate the magnitude of this cross-contamination, we evaluated the amount of overlap by performing separate fits over the $s$- and $p$-shell peaks in the available $^{12}\mathrm{C}$ structure function. We found that for our $p_{\rm miss}$ range, the $p$-shell cut includes around $5\%$ of protons coming from the $s$ shell, whereas the amount of protons coming from the $p$ shell that are included in the $s$-shell cut is negligible. To obtain the corresponding uncertainty, we multiplied these cross-contamination estimates by the relative differences between the individual components for the two shells. Since the difference is positive for one component and negative for the other, we added the uncertainties in quadrature for the single ratio, whereas the uncertainty on the double ratio, although in principle vanishing, is dominated by the $p$-shell single-ratio uncertainty.

The last two items from Table \ref{tab:systematics} correspond to the quality of the spin-precession evaluation in our maximum-likelihood algorithm. We started by comparing the results obtained from employing the spin-transfer matrix to those calculated using the \texttt{QSPIN} program which is more precise but considerably slower. The second contribution arises from the finite resolution of the proton trajectory parameters (e.g.~vertex position). Here we used again \texttt{QSPIN} to evaluate the average dispersion from the analysis of $100$ trajectories with normally distributed variations in each parameter, where its spread was used as the standard deviation of the sampling function. Finally, we obtain the total estimated systematic uncertainty by adding contributions from each source in quadrature. The systematic uncertainties of the polarization components are comparable to the statistical uncertainties.

\section{Results and Discussion}

\begin{figure*}[ht!]
\begin{center}
\includegraphics[width=\textwidth]{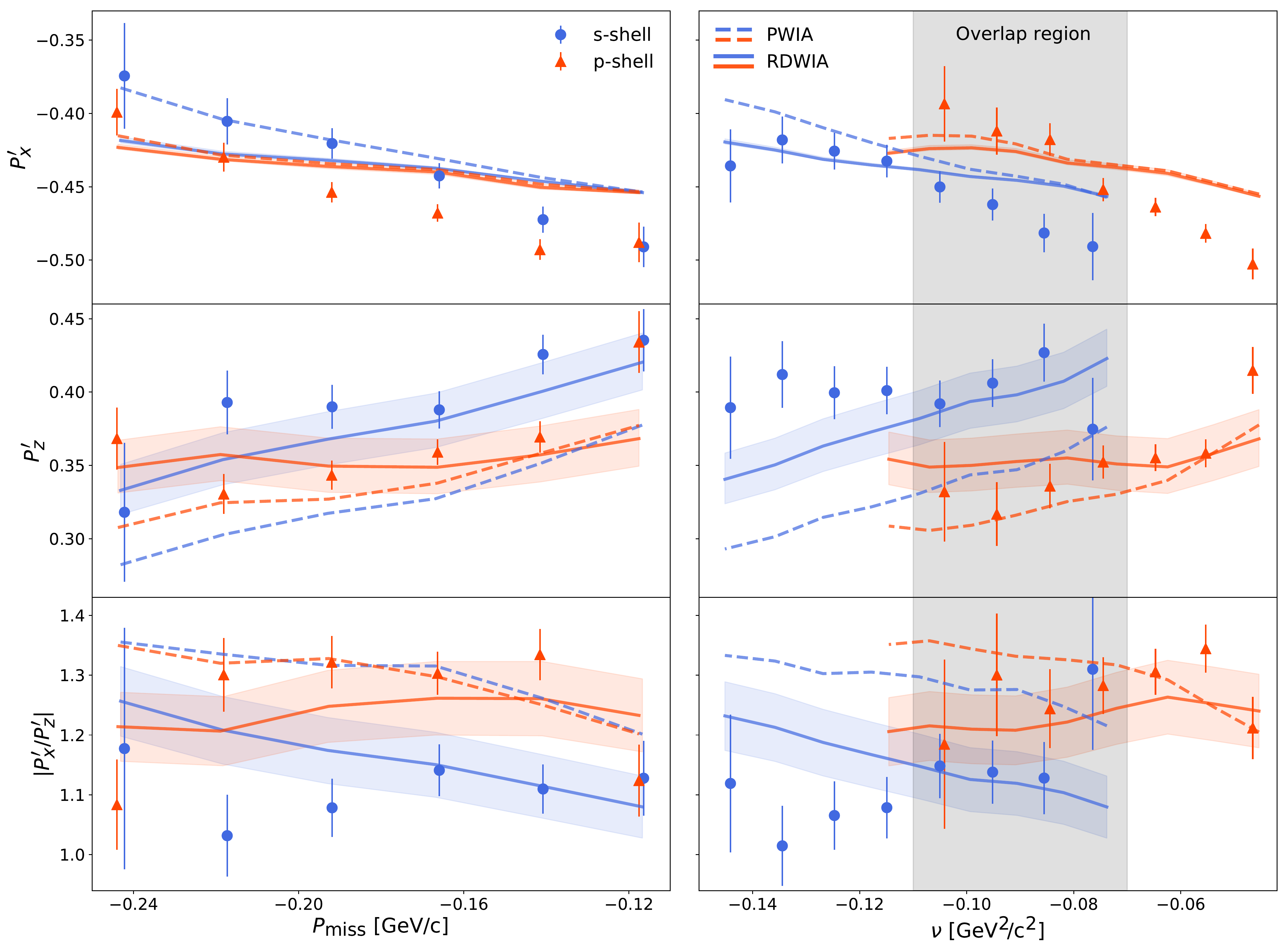}
\caption{
The measured polarization components $P'_x$ (top), $P'_z$ (middle), and their ratio $P'_x/P'_z$ (bottom) as a function of missing momentum (left) and virtuality (right). Shown are statistical uncertainties only. The lines represent RDWIA and PWIA calculations for the corresponding shell obtained using a slightly modified program from \cite{Meucci:2001qc} (see text). The shaded colored regions correspond to RDWIA calculations with the form-factor ratio, $G_{\rm E}/G_{\rm M}$, modified by $\pm 5\%$.   
}
\label{fig:raw_comp_and_R}
\vspace*{\floatsep}
\includegraphics[width=\textwidth]{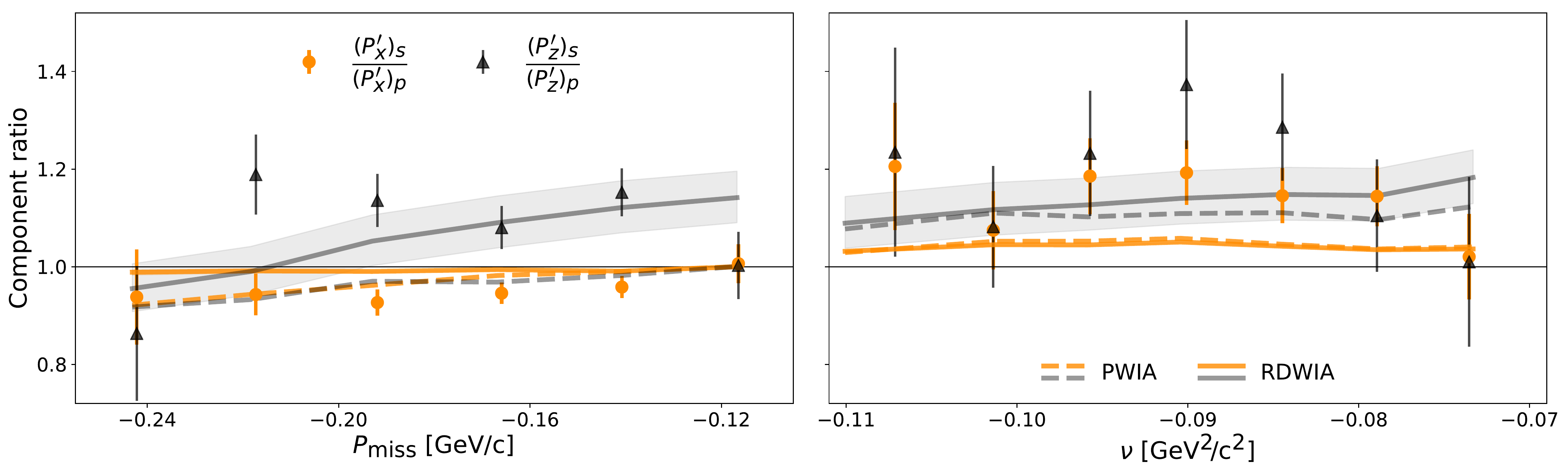}
\caption{
Ratios of given polarization-transfer components ($P^{\prime}_{\!x}$ or $P^{\prime}_{\!z}$) for each shell in $^{12}{\rm C}$ ($s$ or $p$) as a function of missing momentum (left) and virtuality (right). Note that here the virtuality range is narrower since the ratios, which compare the two shells, can be calculated only in the overlap region. The solid and dashed lines represent the RDWIA and the PWIA calculations, respectively. Following the components, only the ratio $(P^{\prime}_{\!z})_s/(P^{\prime}_{\!z})_p$ is sensitive to the electromagnetic form-factor ratio modification, and hence, has a visible band around the calculation. Since we are searching for differences between the two shells, we modified the electromagnetic form-factor ratio only for one of them.   
}
\label{fig:comp_ratio}
\end{center}
\end{figure*}

In the top two panels of Fig. \ref{fig:raw_comp_and_R} we show the polarization-transfer components $P'_x$ and $P'_z$ to protons knocked out from the $s$ and $p$ shells, as a function of the missing momentum and virtuality. Only statistical uncertainties are shown. As in Fig.~\ref{fig:v-pmiss}, the gray band in the right-column plots indicates the virtuality-overlap region between the protons extracted from the $s$ and $p$ shells. The solid lines represent calculations in the relativistic distorted-wave impulse approximation (RDWIA), where we use the average democratic optical potential from \cite{Cooper:2009}, relativistic bound-state wave functions obtained with the NL-SH parametrization \cite{Sharma:1993}, and free-proton electromagnetic form factors from \cite{Bernauer:2014}. Because the original RDWIA program from \cite{Meucci:2001qc} was written for use with coplanar kinematics only, we modified it to include the remainder of the 18 hadronic structure functions present in the ${\rm A}(\vec{e},{e}'\vec{p})$ reaction under the one-photon-exchange approximation \cite{Picklesimer:1989, Boffi:1996}. 

The effects of FSI can be appreciated by comparing the RDWIA (solid lines) and PWIA (dashed lines) calculations showing the different contributions to the transverse and longitudinal components in the two shells. To explore the sensitivity of the polarization components to the ratio $G_{\rm E}/G_{\rm M}$ we repeated the calculation with the form-factor ratio modified by $\pm 5\%$. The impact of this variation on the results of the calculation is shown as a band around the respective calculation with no modification. We note that in this kinematic region, varying the form-factor ratio has a very small effect on the transverse component, $P_{x}^{\prime}$, while the longitudinal component, $P_{z}^{\prime}$, shows a linear dependence on the $G_{\rm E}/G_{\rm M}$, as can be seen in Fig.~\ref{fig:raw_comp_and_R}. The behavior of the individual components is translated to the linear dependence of their ratio, $P'_x /P'_z$, on the form-factor ratio. 

\begin{figure}[th]
\includegraphics[width=.963\columnwidth]{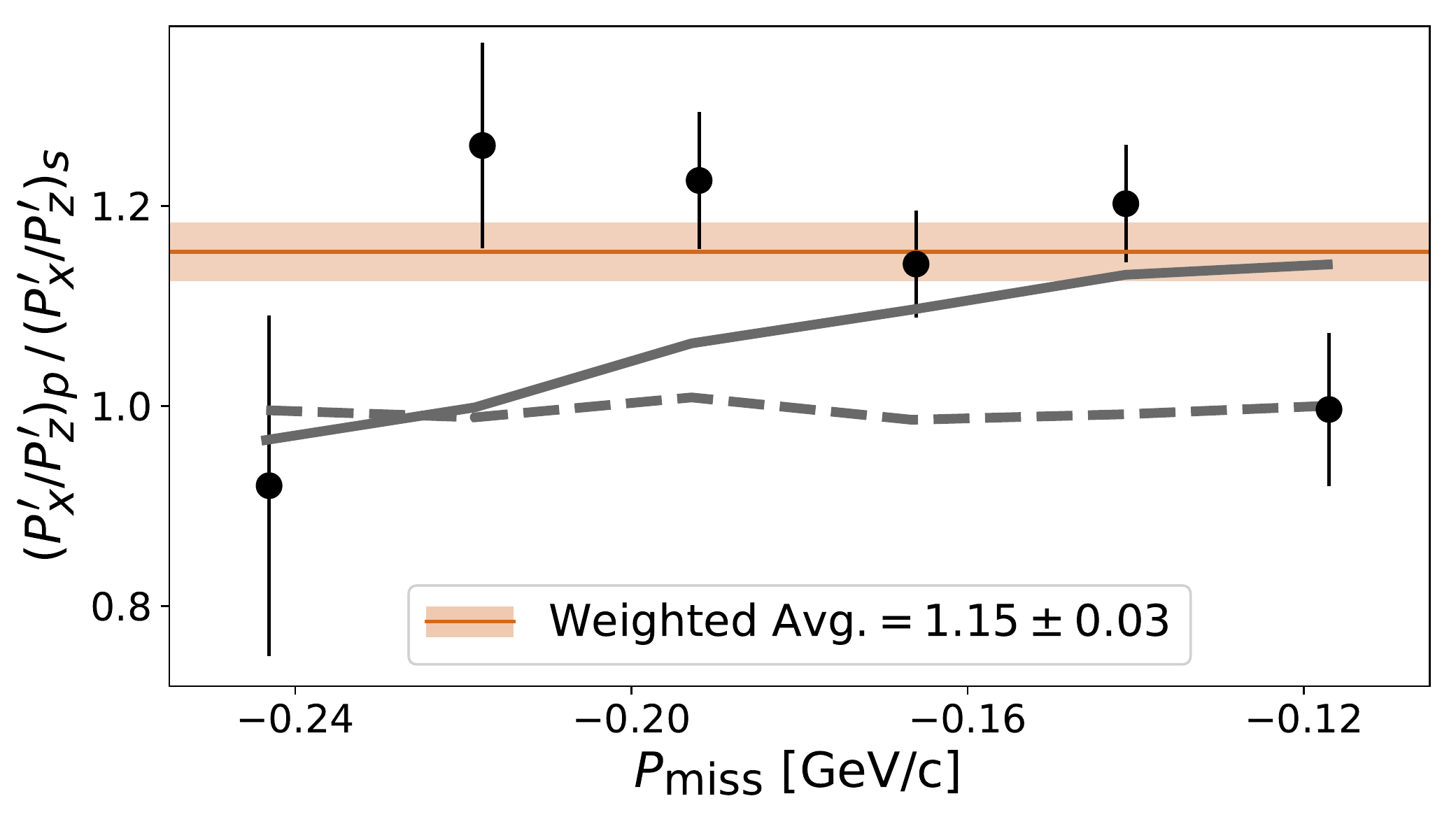}

\includegraphics[width=.975\columnwidth]{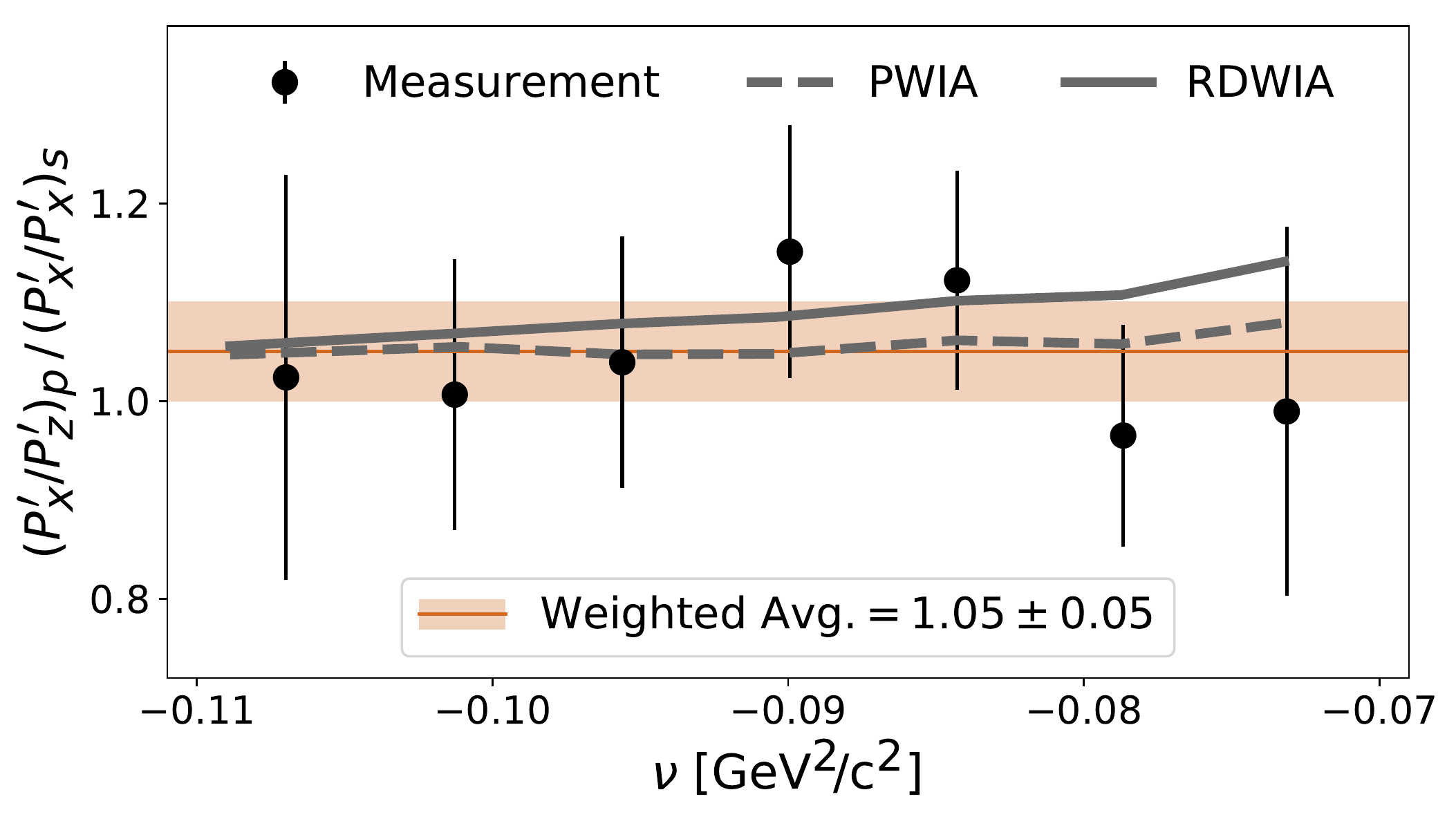}
\caption{ The polarization transfer double ratio as a function of missing momentum (top) and virtuality (bottom). The solid and dashed gray lines represent the RDWIA and the PWIA calculation, respectively, whereas the colored line and band correspond to the weighted average of the measurements and its uncertainty.}
\label{fig:DR}
\end{figure}

Nuclear effects can not only differ for protons from the $s$ and $p$ shells, but may also have different effects on the transverse ($x$) and longitudinal ($z$) polarization components when we consider the protons from a single shell. This can be seen as a deviation from unity in the bottom panel of Fig. \ref{fig:raw_comp_and_R}, where we show $P_x' /P_z'$ for each shell separately, as well as in Fig. \ref{fig:comp_ratio}, which includes component ratios, $P_{i}^{\prime s} /P_i^{\prime p}$ ($i=x,z$) for the two shells. Such differences are also foreseen by the theoretical calculations. Furthermore, the deviations between the calculations and the data do not exclude medium modifications like those observed in the unpolarized $(e,e'p)$ \cite{Steenhoven:1986} and polarization transfer measurements \cite{Brecelj:2020}, which are not accounted for in the calculations. Searching for the \textit{density-dependent} in-medium modifications we examine the double ratio $(P_x^{\prime} /P_z^{\prime} )_p/(P_x^{\prime} /P_z^{\prime})_s$ to minimize the contributions of those effects. The double ratio is shown for the measured components as a function of the missing momentum in the top panel of Fig. \ref{fig:DR} along with the calculated double ratio under RDWIA (solid line) and PWIA (dashed line). One expects this double ratio to be unity if nuclear effects cancel and there are no modifications in the form-factor ratios. The measured double ratio is not unity but it is almost constant with a weighted average of $1.15\pm 0.03$, indicating that FSI contributions do not cancel in the double ratio. This deviation from unity is predicted by the RDWIA calculations (solid line), while the PWIA (dashed line) is consistent with unity. This indicates that the deviation is due to nuclear effects, which do not cancel in this comparison. 

It has been shown \cite{Yaron:2017, IzraeliYaron:2018, Paul:2019} that virtuality is a useful parameter to compare the polarization transfer to protons bound in different nuclei. The data from the two shells in the overlap region (where the knockout protons from the two shells can be compared at the same $\nu$) are shown as a function of the proton virtuality in the bottom panel of Fig. \ref{fig:DR}. The weighted average of $1.05\pm 0.05$ is consistent with unity, as predicted by both RDWIA and PWIA calculations. We note that the nuclear effects are much reduced in the RDWIA-calculated double ratio (close to the PWIA prediction), supporting the comparison at the same virtuality. Even though the polarization transfer observables are modified by the medium, the measured double ratio for the $s$- and $p$-shell protons being consistent with unity suggests that no density-dependent differences in medium modification of the proton (e.g. its form-factor ratio) are observed. The small deviation of the double ratio from unity can already be accounted for with the unmodified electromagnetic form-factor ratio and simple PWIA calculations, while the measurements are also in agreement with the RDWIA calculations (reduced $\chi^2 = 0.48$, $p=0.89$). Thus, we found no evidence of density-dependent modifications. 

The ratios $P_x^{\prime}/P_z^{\prime}$ of components of polarization transfer 
to deeply bound protons were measured for several nuclei. It was shown \cite{Yaron:2017, Izraeli:2018, Paul:2019} that a comparison of this ratio to that of a free proton, $(P_x^{\prime} /P_z^{\prime} )_A /(P_x^{\prime} /P_z^{\prime} )_H$, at given $\nu$ shows the same deviations for $^2 {\rm H}$, $^4 {\rm He}$, and $^{12} {\rm C}$ despite different kinematic conditions. The agreement of the results when the proton is bound in $^{2}\mathrm{H}$, which is a slightly-bound two-body system and often used as an effective neutron target, with those obtained in nuclei with a high average nuclear density (like $^{4}\mathrm{He}$ and $^{12}\mathrm{C}$) also supports our observation. While FSI and the local nuclear density may differ between these nuclei, their effect on the polarization transfer is similar, and no nuclear-density-dependent modifications are observed.

\FloatBarrier

\section{Conclusions}
We presented measurements of the polarization transfer to deeply bound protons
in the $s$ and $p$ shells of ${}^{12}\mathrm{C}$ by polarized electrons with the $^{12}\mathrm{C}(\vec{e},e'\vec{p}\,)$ reaction. To investigate the nuclear-density dependence and possible in-medium modification of the proton electromagnetic form factors, we exploited the fact that the ratio of the transverse to longitudinal components is sensitive to the electromagnetic form-factor ratio. The measured polarization ratios for protons extracted from the two shells were studied and compared as a function of either the missing momentum or the bound-proton virtuality. Although according to some theories there is a large difference in the nuclear density between the two shells in ${}^{12}\mathrm{C}$, the measurements show no significant differences between the $s$- and $p$-shell protons to the level of 5\% when compared at the same virtuality, as expected from both PWIA and RDWIA calculations. One cannot exclude density-dependent effects, which may conspire to cancel and hide a density-dependent medium modification of the form factors. However, the straightforward explanation of the measurement reported here is that there are no differences in medium
modifications of the proton form-factor ratio when comparing $s$- and $p$-shell protons in $^{12}\mathrm{C}$.

\section{Acknowledgments}

We would like to thank the Mainz Microtron operators and technical crew for the excellent operation of the accelerator. This work is supported by the Israel Science Foundation (Grants 390/15, 951/19) of the Israel Academy of Arts and Sciences, by the PAZY  Foundation (Grant 294/18), 
by the Israel Ministry of Science, Technology and Space, by the Deutsche Forschungsgemeinschaft (Collaborative Research Center 1044), by the U.S. National Science Foundation (PHY-1205782).  
We acknowledge the financial support from the Slovenian Research Agency (research core funding No. P1-0102) and from the Croatian Science Foundation (under the project 8570).


\bibliography{ceep2017}
\end{document}